\documentclass[
reprint,
superscriptaddress,
amsmath,amssymb,
aps,
pra,
noeprint
]{revtex4-2}
 
\usepackage{amsmath}
\usepackage{amssymb}
\usepackage{amsfonts}
\usepackage{booktabs}
\usepackage{dsfont}
\usepackage{graphicx}
\usepackage{hyperref}
\usepackage{multirow}
\usepackage[utf8]{inputenc}
\usepackage[T1]{fontenc}
\usepackage[english]{babel}
\usepackage{xcolor}
\usepackage{tikz}
\usepackage[normalem]{ulem}
\graphicspath{{../src/allowed_areas_July/}{../figures/}}

\newcommand{\avg}[1]{\left\langle #1 \right\rangle}
\newcommand{\bigexp}[1]{\exp{\left\{ #1 \right\}}}
\newcommand{\ha}{\hat{a}}
\newcommand{\Oeff}{\Omega_\mathrm{eff}}
\newcommand{\Ums}{\hat{U}_\mathrm{MS}}
\newcommand{\MS}{\mathrm{MS}}
\newcommand{\ph}{\mathrm{ph}}
\newcommand{\Ca}{{}^{40}\mathrm{Ca}^+}
\newcommand{\Rxx}{R_\mathrm{XX}}

\let\Re\relax
\DeclareMathOperator{\Re}{Re}
\let\Im\relax
\DeclareMathOperator{\Im}{Im}
\DeclareMathOperator{\Tr}{Tr}

\begin{document}
\title{Fast Mølmer-Sørensen gates in trapped-ion quantum processors with compensated carrier transition}
\date{\today}

\author{Evgeny Anikin}
\email[]{evgenii.anikin@skoltech.ru}
\affiliation{Russian Quantum Center, Skolkovo, Moscow 143025, Russia}

\author{Andrey Chuchalin}
\affiliation{Russian Quantum Center, Skolkovo, Moscow 143025, Russia}
\affiliation{Moscow Institute of Physics and Technology, Dolgoprudny, 141700, Russia}

\author{Nikita Morozov}
\affiliation{Russian Quantum Center, Skolkovo, Moscow 143025, Russia}

\author{Olga Lakhmanskaya}
\affiliation{Russian Quantum Center, Skolkovo, Moscow 143025, Russia}

\author{Kirill Lakhmanskiy}
\affiliation{Russian Quantum Center, Skolkovo, Moscow 143025, Russia}
\begin{abstract}
  Carrier transition is one of the major factors hindering the high-speed implementation 
  of the M{\o}lmer-S{\o}rensen gates in trapped-ion quantum processors. 
  We present an approach to design laser pulse shapes for the M{\o}lmer-S{\o}rensen gate in ion chains which accounts for the effect of carrier transition on 
  qubit-phonon dynamics.  We show that the fast-oscillating carrier term effectively modifies the spin-dependent forces acting on ions, and this can be 
  compensated by a simple nonlinear transformation of a laser pulse. Using numerical simulations for short ion chains and perturbation theory for longer chains up to $20$ ions, 
  we demonstrate that our approach allows to suppress the infidelity contribution from off-resonant carrier excitation from $10^{-3}$-$10^{-2}$ to values below $10^{-4}$, whereas 
  the gate duration remains of the order of tens of microseconds.
\end{abstract}
\maketitle

\section{Introduction}

High-fidelity entangling gates are critical for the practical realization of quantum computers. 
With cold trapped ions, remarkable progress in such gate implementation has been achieved in recent decades \cite{Bruzewicz2019}. 
Since the Cirac-Zoller gate \cite{Cirac1995}, various gate designs have been proposed and implemented, including M{\o}lmer-S{\o}rensen gate \cite{Soerensen2000}, 
light-shift gate \cite{Leibfried2003a}, ultrafast gates \cite{Mizrahi2013a}, etc.
Among them, the M{\o}lmer-S{\o}rensen gate and its variations are the most ubiquitous.

The M{\o}lmer-S{\o}rensen gate MS gate is implemented with the help of a bichromatic external field pulse detuned symmetrically 
from the qubit transition, typically optical or Raman \cite{Soerensen2000}.
In the Lamb-Dicke regime, the field creates a spin-dependent force acting on ions, and the pulse parameters should be chosen to make ion trajectories closed 
curves in phase space. As a result, qubits acquire a 
spin-dependent phase and become entangled. 

For a two-ion crystal, high-fidelity MS gates can be implemented with constant-amplitude field smoothly turned on and 
off \cite{Kirchmair2008}. For larger ion number, more complicated pulse shapes are required in order to close phase
space trajectories for multiple motional modes of an ion crystal \cite{Choi2014}. For amplitude-modulated pulses, the 
pulse shape is determined by a set of linear equations \cite{Choi2014} which can be efficiently solved numerically.

In recent literature, a variety of pulse shaping approaches have been proposed for different  
purposes. Apart from amplitude modulation proposed in \cite{Choi2014}, phase modulation \cite{Milne2020} and 
multi-tone beams \cite{Shapira2018} can be used. Also, additional constraints on the pulse shape allow to 
implemement parallel gates \cite{Grzesiak2020, Olsacher2020}, to enhance gate robustness 
against fluctuations of the experimental 
parameters \cite{Bluemel2021, Ruzic2024}, or to minimize the total external field power \cite{Bluemel2021}. 

A careful error analysis is necessary to design laser pulses implementing 
MS gate with highest possible fidelity.
The errors originate both from technical noises, such as laser and magnetic field fluctuations, and the 
intrinsic laser-ion interactions \cite{Wu2018}. To account for the latter, it is necessary to go beyond the 
usual simplifying assumptions such as Lamb-Dicke and rotating-wave approximations often used to describe laser-ion dynamics.
The latter can be achieved by a numerical simulation in the full ion-phonon Hilbert space
or by applying a systematical procedure such as Magnus expansion beyond the leading order \cite{Vedaie2023, Bluemel2024}.

One of the unwanted interactions affecting the trapped-ion dynamics during the gate operation is the carrier transition \cite{Wu2018}.  
For MS gate operation, the bichromatic beam components should be tuned closely to qubit motional sidebands. 
However, due to finite gate duration, the off-resonant direct (carrier) 
transition between qubit levels is also present. 
Its influence on the gate dynamics depends significantly 
on the type of the laser beam configuration. 
For example, the recent implementation of the MS gate with a phase-stable standing wave 
\cite{Saner2023} allows to eliminate carrier transition almost entirely. Also, for  
Raman qubits in phase-insensitive geometry \cite{Lee2005}, carrier transition causes only an 
additional spin-dependent phase. However, carrier transition significantly contributes into 
gate dynamics for two widely used beam configurations:
for a single bichromatic beam with two co-propagating components 
and for Raman qubits in phase-sensitive geometry.
In these cases carrier transition adds as a non-commuting term to the 
spin-dependent force Hamiltonian. Although its presence can be utilized 
to create new types of interactions between ion qubits \cite{Bazavan2023}, 
carrier term becomes one of the dominant factors limiting gate fidelity at short gate durations. 

The theoretical analysis of ion dynamics with full account of non-commuting carrier term 
has been performed for two ions interacting with a single phonon mode \cite{Roos2007, Kirchmair2008, Bazavan2023}. 
It has been shown that the fast-oscillating carrier term results in a nonlinear 
renormalization of the driving field amplitude, and (for strictly rectangular pulse) in a basis rotation of the 
$\Rxx(\theta)$ gate operator depending on the relative phase between the bichromatic beam components. 
However, such analysis has not been performed for longer ion chains where the dynamics of multiple phonon modes should be considered.

In this work, we present a theoretical analysis of the influence of carrier transition on 
MS gate dynamics in an ion chain with account of all phonon modes. Our analysis applies to a bichromatic beam with 
two co-propagating components, which corresponds to the cases of optical qubits and Raman qubits in phase-sensitive geometry.
For amplitude-shaped pulse, we eliminate the carrier term in the system Hamiltonian by transitioning into the interaction picture. 
The leading part of the resulting Hamiltonian has the form of a spin-dependent force Hamiltonian, 
where carrier transition modifies the force values obtained from Lamb-Dicke expansion. 
We propose a family of pulse shapes which take this modification into account. Our pulses 
can be found by applying a simple nonlinear transformation to a pulse obtained from linear equations. 
We use numerical simulations for short chains and perturbation theory calculations for longer chains to demonstrate
that our pulse shaping scheme shows a considerable fidelity gain in comparison to the pulses obtained from linear equations.
In particular, we demonstrate the theoretical infidelity below $10^{-4}$ for our pulses for 2-20 ions in the chain.

So, our pulse shaping approach provides a considerable speedup while maintaining 
high gate fidelity, which is an important step towards meeting the challenging 
requirements for practical quantum computations.

\section{Dynamics of trapped ions qubits in the presence of carrier transition}
\label{sec:evo_op_w_carrier}
We consider a trapped-ion quantum processor consisting of a system of ions 
confined in a radio frequency Paul trap irradiated by laser light. Two electronic levels 
of each ion constitute the qubit levels. We assume that the ions form a linear crystal, therefore,
ions share independent sets of collective phonon modes for each trap axis direction \cite{James1998}.
We consider the implementation of an entangling gate between two ion qubits,
namely M{\o}lmer-S{\o}rensen gate (MS gate) \cite{Soerensen2000}. For that,
bichromatic laser field symmetrically detuned from qubit transition is required, and we focus on the case of 
amplitude-modulated field. The interaction-picture Hamiltonian for the considered pair of ions reads \cite{Monroe2021}
\begin{equation}
	\begin{gathered}
  \hat{H} = -i\sum_{i=1,2} \Omega(t)\cos{(\mu t + \psi)}\left( e^{ik_a\hat{r}_{ia}} \sigma^i_+  - \mathrm{h.c.}\right),\\
     k_a\hat{r}_{ia} = \sum_{m=1}^{n_\mathrm{ions}}\eta_{im}(\hat{a}_m e^{-i\omega_m t} + \hat{a}_m^\dagger e^{i\omega_m t}),
  	\end{gathered}
  \label{eq:full_ham}
\end{equation}
where $k_a$ are the components of the laser field wavevector, $\hat{r}_{ia}$ are the components of ions displacememnts from
their equilibrium positions,
$\hat{a}_m^\dagger$ and $\hat{a}_m$ are the creation and annihilation operators $\hat{a}_m, \hat{a}_m^\dagger$ 
of the phonon modes, $\omega_m$ are the phonon mode frequencies, $\eta_{im}$ are the Lamb-Dicke parameters, 
$\Omega(t)$ is the bichromatic beam amplitude envelope, $\mu$ is the detuning between the bichromatic beam components, 
and $\psi$ equals half the phase difference between the components of the bichromatic beam.
The action of our target MS gate operator on ion qubits is represented by an operator
$\Rxx(\phi) = \bigexp{-i\phi\sigma_x\otimes \sigma_x}$, where $\phi$ denotes the spin-spin coupling phase. 
For gate implementation,
one should find a pulse shape $\Omega(t)$ so that the system evolution operator reduces to the target gate operator. 

Typically, the approximate evolution operator is found from the expansion of the system Hamiltonian in the 
Lamb-Dicke parameters \cite{Monroe2021}. Up to the first order in the expansion, the Hamiltonian reads
\begin{equation}
  \hat{H} \approx \hat{H}_\mathrm{LD} = \hat{H}_0 + \hat{H}_1,
  \label{eq:ld_ham}
\end{equation}
where
\begin{equation}
 \hat{H}_0 = \sum_{i=1,2} \Omega(t)\cos{(\mu t + \psi)} \sigma_{y}^i.
\end{equation}
\begin{multline}
  \hat{H}_1 = \sum_{i=1,2}\sum_{m=1}^{n_\mathrm{ions}} \eta_{im}\Omega(t)\cos{(\mu t + \psi)}\times\\
  \times(\hat{a}_m e^{-i\omega_m t} + \hat{a}_m^\dagger e^{i\omega_m t})  \sigma_{x}^i
  \label{eq:pure_ms}
\end{multline}

The Hamiltonian \eqref{eq:ld_ham} in the Lamb-Dicke approximation (LD Hamiltonian)
contains two terms. The term $\hat{H}_1$ is a
spin-dependent force Hamiltoninan \cite{Haljan2005, Zhu2006a}. 
The term $\hat{H}_0$ (carrier term) corresponds to direct carrier transitions. 
The contribution of the carrier term is often neglected. This approximation 
is valid when $\mu$ is close to one or several phonon mode frequencies, so 
$\hat{H}_1$ contains slowly-oscillating terms which 
contribute the dynamics significantly, whereas $\hat{H}_0$ 
is fastly oscillating. Further, we keep both terms and analyze the contribution of $\hat{H}_0$ to gate dynamics.

The evolution operator for the spin-dependent force Hamiltonian $\hat{H}_1$ reads \cite{Zhu2006a} 
\begin{multline}
  \hat{U}_\MS(t_1, t_2) = 
  \exp\left(-\frac{i}{2}\sum_{i,j} \chi_{i,j}^0(t_2, t_1) \sigma_{x}^i \sigma_{x}^j\right)\\
    \prod_m D_m\left(\sum_i\sigma^i_{x}\alpha^0_{i,m}(t_2, t_1))\right),
    \label{eq:propagator_nocar}
  \end{multline}
where 
\begin{equation}
  \alpha_{im}^0(t_2, t_1) = -i\int_{t_1}^{t_2} f^0_{im}(t') dt' 
  \label{eq:alpha_no_carrier}
\end{equation}
\begin{equation}
\chi_{ij}^0(t_2, t_1) = 2\Re\int_{t_1}^{t_2} 
\alpha_{im}^0(t', t_1) (f^0_{jm})^*(t') \,dt',
  \label{eq:chi_no_carrier}
\end{equation}
\begin{equation}
  f^0_{im}(t) = \eta_{im}e^{i\omega_m t}\Omega(t)\cos{(\mu t + \psi)}.
  \label{eq:f0_0}
\end{equation}

An appropriate pulse shape $\Omega(t)$ should be found for the implementation of an $\Rxx(\phi)$ gate.
Let $\Omega(t)$ be applied in the time interval $(t_0, t_f)$. 
The evolution operator \eqref{eq:propagator_nocar} reduces to the target $\Rxx(\phi)$ operator provided that
the following conditions for 
the displacement amplitudes
$\alpha^0_{im}$ and 
the spin coupling phases
$\chi^0_{ij}$ are satisfied \cite{Zhu2006a, Choi2014}:
\begin{equation}
  \alpha_{im}^0(t_f, t_0) = 0,\\
    \label{eq:gate_condition_alpha_nocar}
\end{equation}
\begin{equation}
  \chi_{12}^0(t_f, t_0) = \phi. 
    \label{eq:gate_condition_chi_nocar}
\end{equation}
The equations \eqref{eq:gate_condition_alpha_nocar} comprise a homogeneous system of 
$2n_\mathrm{ions}$ linear equations for $\Omega(t)$. 
They can be solved numerically by standard linear algebra routines and define $\Omega(t)$ up to a normalization constant. The latter can be found from the quadratic equation \eqref{eq:gate_condition_chi_nocar}. 
With approximations made, the resulting 
$\Omega(t)$ implements the MS gate with 100\% fidelity.

However, the contribution of the carrier term $\hat{H}_0$ into infidelity 
grows with increasing gate speed even in the
absence of technical noises.
To account for the carrier term systematically, let us switch into the interaction picture with 
respect to $H_0$:
\begin{equation}
  |\psi\rangle = e^{-i\int_{t_i}^{t}\hat{H}_0(t') dt'} |\psi_{Ic}\rangle
  = e^{-i\Phi(t)\sum_i \sigma^i_{y}}|\psi_{Ic}\rangle,
  \label{eq:int_transform}
\end{equation}
where
\begin{equation}
  \Phi(t) = \int_{t_0}^{t} dt' \Omega(t') \cos{(\mu t' + \psi)}.
  \label{eq:phi}
\end{equation}

After this transformation the Hamiltonian takes the form 
 \begin{multline}
   \hat{H}_{Ic} = \hat{H}_{Ic}^{(0)} + \hat{V} = \\
   =\sum_{i,m} \eta_{im}\Omega(t)\cos{(\mu t + \psi)}(\hat{a}_m e^{-i\omega_m t} + \hat{a}_m^\dagger e^{i\omega_m t})\times\\
  \times(\underbrace{\cos{2\Phi(t)}\sigma_{x}^i}_\text{leading-order} + \underbrace{\sin{2\Phi(t)} \sigma_z^i}_\text{perturbation}).
  \label{eq:int_ham}
\end{multline}
The rotation angle $\Phi(t_f)$ entering the transformation given by Eq.~\eqref{eq:int_transform} can be greatly reduced by 
choosing such pulse $\Omega(t)$ that
satisfies the following
smoothness conditions:
\begin{enumerate}
  \item $\Omega(t)$ varies slowly in comparison to $\mu^{-1}$,
  \item $\Omega(t)$ and its first derivatives vanish at the beginning 
    and at the end of the pulse.
\end{enumerate}
Under these conditions and with our proposed pulse design (described in more detail in
subsequent sections), we get typical values of $\Phi(t_f)$ of the  order of 
$10^{-5}$-$10^{-4}$. Therefore, we can assume that the final state of the 
qubit-phonon system is determined only by evolution with the Hamiltonian
\eqref{eq:int_ham}.

For further analysis of the Hamiltonian \eqref{eq:int_ham}, 
it is convenient to split it into 
two parts, the leading order part $\hat{H}_{Ic}^{(0)}$ and 
the perturbation part $\hat{V}$, as shown by underbraces in Eq.~\eqref{eq:int_ham}. 
Such a decomposition is justified under 
two assumptions on $\Phi(t)$:
\begin{enumerate}
  \item $\Phi(t)$ oscillates near zero;
  \item its magnitude remains $\lesssim 1$.
\end{enumerate}
These assumptions hold for all the pulses that will be considered below. 
With them, the $\cos{2\Phi(t)}$ term oscillates near 
$\sim 1$ and gives a contribution important on the large timescales. In contrast,
$\sin{2\Phi(t)}$ oscillates near zero, so its contribution cancels on the large timescales. Because of that, it is reasonable to take the sine term as a
perturbation, where $\Phi(t)$ is considered as a small parameter.

The leading-order term $\hat{H}^{(0)}_{Ic}$ has the form of the spin-dependent forces Hamiltonian, so its 
evolution operator has the same form as \eqref{eq:propagator_nocar}:
\begin{multline}
  \hat{U}_0(t_1, t_2) = 
    \exp\left(-\frac{i}{2}\sum_{i,j} \chi_{i,j}(t_2, t_1) \sigma_{x}^i \sigma_{x}^j\right)\\
    \prod_m D_m\left(\sum_i\sigma^i_x\alpha_{i,m}(t_2, t_1)\right),
    \label{eq:propagator}
  \end{multline}
where 
\begin{equation}
  \alpha_{im}(t_2, t_1) = -i\int_{t_1}^{t_2} f_{im}(t') dt',
  \label{eq:alpha_w_carrier}
\end{equation}
\begin{equation}
\chi_{ij}(t_2, t_1) = 2\Re\int_{t_1}^{t_2} 
\alpha_{im} f_{jm}^* 
dt',
  \label{eq:chi_w_carrier}
\end{equation}
\begin{equation}
  f_{im}(t) = f^0_{im}(t)\underbrace{\cos{2\Phi(t)}}_\text{carrier effect}.
  \label{eq:f}
\end{equation}
Additional  $\cos{2\Phi(t)}$ term in $f_{im}(t)$ arises 
because of the transformation into the interaction picture generated by the carrier term. 

The perturbation term $\hat{V}$ generates corrections to the evolution operator. 
Within the first-order perturbation theory,
the evolution operator of the Hamiltonian \eqref{eq:int_ham} can 
be expressed as 
\begin{equation}
  \hat{U} = \hat{U}_0 \left(\mathds{1} - i\hat{T}_1\right),
  \label{eq:pert_first_ord_U}
\end{equation}
where 
\begin{equation}
  \hat{T}_1 = \int_{t_0}^{t_f} \hat{U}_0^\dagger(t', t_0) \hat{V} 
  \hat{U}_0(t', t_0) dt'.
  \label{eq:pert_first_ord_T}
\end{equation}
As the perturbation $\hat{V}$ contains the sigma matrices $\sigma_{z}^i$,
the operator $\hat{T}_1$ causes the spin flip processes in
$x$-basis: for the initial qubit state $|s_1s_2\rangle_x$ (Pauli string in $x$-basis), 
it causes transitions to another Pauli string $|s'_1s'_2\rangle_x$ which differs from the
initial state by a single spin flip.
In the next sections, we show that the contribution of $\hat{T}_1$ into gate dynamics is 
quite small.
Therefore, in the next sections, we use Eq.~\eqref{eq:propagator} for pulse shape design, 
and Eqs.~\eqref{eq:pert_first_ord_U}, \eqref{eq:pert_first_ord_T} are used only to calculate 
contributions to gate infidelity.

With modified expressions for spin-dependent forces, we get 
new conditions for the implementation of $\Rxx(\phi)$:

\begin{equation}
  \begin{gathered}
    \alpha_{im}(t_f, t_0) = 0,\\
    \chi_{12}(t_f, t_0) = \phi.
  \end{gathered}
  \label{eq:gate_conditions}
\end{equation}

The definitions of $\alpha_{im}$ and $\chi_{ij}$ differ from 
$\alpha_{im}^{0}$ and $\chi_{ij}^{0}$ by the 
the $\cos{2\Phi(t)}$ term, where $\Phi(t)$ is expressed as an integral \eqref{eq:phi} 
containing $\Omega(t)$. Because of that, the Eqs.~\eqref{eq:gate_conditions} 
are nonlinear integral equations 
in contrast to the linear equations \eqref{eq:gate_condition_alpha_nocar}.
In Section~\eqref{sec:pulse_shaping}, we present a scheme for approximate solution
of these equations.

\section{Contributions of carrier term to gate infidelity}
\label{sec:carrier_infidelity}
The expressions \eqref{eq:propagator} and \eqref{eq:pert_first_ord_U} 
for the propagator can be used for analytical calculation of the gate fidelity 
of an ion chain of arbitrary length. In this section, we present the expressions for 
the gate fidelity defined in the full ion-phonon Hilbert space. 
We assume that the ion chain is cooled to the ground state at the beginning of the
gate operation, so the initial state of the qubit-phonon system is 
$|\psi_0\rangle \equiv |\psi_{0q}\rangle\otimes|0_\mathrm{ph}\rangle$,
where  $|\psi_{0q}\rangle$ is the initial qubit state.
The action of the ideal $\Rxx(\phi)$ gate turns qubits
into the state $\Rxx(\phi)|\psi_{0q}\rangle$ but leaves all phonon modes 
in the ground state. Therefore, we can characterize the gate realization by 
fidelity between the target state 
$|\psi_t\rangle \equiv \Rxx(\phi)|\psi_{0}\rangle$ 
(the outcome of the ideal $\Rxx(\phi)$ gate) and the state 
$|\psi(t_f)\rangle = \hat{U}|\psi_0\rangle$ obtained after the evolution with 
the Hamiltonian \eqref{eq:full_ham}:
\begin{equation}
  F_\mathrm{tot} = |\langle\psi_t |\psi(t_f)\rangle|^2 = 
  |\langle \psi_0|\Rxx(\phi)^\dagger \hat{U} |\psi_0\rangle|^2.
\label{eq:fidelity_def}
\end{equation}
This fidelity definition differs from that of  \cite{Wu2018}, \cite{Zhu2006a} as we do not trace 
out phonon degrees of freedom. Further, we will analyze different contributions into the total infidelity arising from the presence of the carrier term.

First, we define the fidelity measure
$F_0$ (zero-order fidelity) which quantifies the differencde of the leading-order propagator $\hat{U}_0$ from
the target gate:

\begin{equation}
  F_0 \equiv |\langle \psi_0 | \Rxx^\dagger(\phi) \hat{U}_0 |\psi_0\rangle|^2
  \label{eq:F0_def}
\end{equation}
In Appendix~\ref{appendix:fidelity_expr}, 
we find $F_0$ for various initial states $|\psi_{0q}\rangle$.
In particular, for 
$|\psi_{0q}\rangle = |s_1s_2\rangle_z$ (Pauli strings in the $z$-basis, $s_i = \pm 1$), the infidelity reads
\begin{equation}
  1-F_0  \approx \sum_{im} |\alpha_{im}(t_f)|^2 + \Delta\chi_{12}(t_f)^2,
  \label{eq:inf0_z}
\end{equation}
where $\Delta\chi_{12}(t_f) = \phi - \chi_{12}(t_f)$ is the error in the 
spin coupling phase.
We see that incomplete closure of the phase trajectories and the error in the spin coupling phase give two additive 
contributions into the average gate infidelity. 

For $|\psi_{0q}\rangle = |s_1s_2\rangle_x$ (Pauli strings in $x$-basis), 
zero-order gate infidelity is given by expression
\begin{equation}
  \begin{gathered}
    1 - F_0 = P_\mathrm{ph}^s = \sum_m \left|\sum_{i} \alpha_{im} s_i\right|^2.
  \end{gathered}
  \label{eq:inf_x0}
\end{equation}
In this case, the error in $\phi$ does not contribute to gate fidelity. At small $\alpha_{im}$, the infidelity
can be interpreted as the probability of phonon excitation after the gate operation.

Second, we determine the contribution to infidelity $P^s_\mathrm{flip}$ caused by the spin-flip
perturbation accounted by the $\hat{T}_1$ term in the 
evolution operator \eqref{eq:pert_first_ord_U}. For simplicity, we 
consider only the initial states $|\psi_0\rangle = |s_1s_2\rangle_x$.
The spin flip probability for such initial states is
\begin{equation}
  P^s_\mathrm{flip} = \langle s, 0_\ph | \hat{T}_1^\dagger \hat{T}_1 | s, 0_\ph\rangle.
  \label{eq:spin_flip_error}
\end{equation}
In Appendices~\ref{appendix:fidelity_expr} and \ref{appendix:spin_flip}, we derive the
explicit expression \eqref{eq:spin_flip_error_explicit} for $P^s_\mathrm{flip}$ as a two-dimensional integral over time. 
(The Eq.~\eqref{eq:spin_flip_error_explicit} is too long to present it in the main text.)
The integral \eqref{eq:spin_flip_error_explicit} is suitable for the calculation of the spin flip probability for long ion 
chains where the full numerical solution is not feasible. Also, we show that the gate infidelity for the initial states $|s_1s_2\rangle_x$ with account of the carrier transition
reads
\begin{equation}
  \begin{gathered}
    1 - F_\mathrm{tot} = P_\mathrm{ph}^s + P^s_\mathrm{flip},\\
  \end{gathered}
  \label{eq:inf_x}
\end{equation}
so the contributions of imperfectly closed phase trajectories and spin flip error are additive. For other qubit states,
there are interference terms between the two considered contributions.

Finally, we derive an upper bound for the infidelity averaged over all initial qubit states over the Fubini-Study measure \cite{Bengtsson2006}:
\begin{equation}
  1-\avg{F_\mathrm{tot}} < 1-\avg{F_0} + \frac{1}{4}\sum_s P^s_\mathrm{flip}, 
\end{equation}
where $F_0$ given by Eq.~\eqref{eq:F0_def} accounts only for the leading-order term of $\hat{U}$. Therefore, it is sufficient
to calculate the contribution of $\hat{T}_1$ only for the initial states $|s_1s_2\rangle_x$ to find an upper bound for 
the average gate infidelity.

Thus, carrier term contributes into two types of gate error. First, it modifies the values of the spin-dependent forces,
so it causes incomplete closing of the phase trajectories unless the pulse shape is chosen appropriately. Second, it 
causes the spin-flip error arising from $\hat{T}_1$. 
In the following, 
we present a method to find pulse shapes which
minimize the error of the first kind.

\begin{figure}[h]
  \centering
  \begin{minipage}{\linewidth}
    \includegraphics[width=\linewidth]{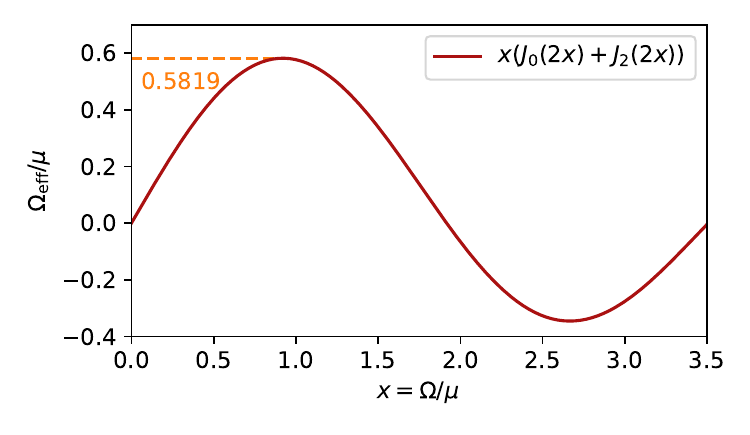}
  \end{minipage}
  \caption{(a) The dependence of the effective Rabi frequency $\Omega_\mathrm{eff}(t)$ on $\Omega$ given by Eq.~\eqref{eq:eff_omega}.}
  \label{fig:omega_eff}
\end{figure}

\section{Compensation of carrier error by a transformation of an amplitude-shaped pulse}
\label{sec:pulse_shaping}
In this section, we present a method to find pulse shapes 
as approximate solutions of the Eqs.~\eqref{eq:gate_conditions}. 
This compensates the modification of the spin-dependent forces by the carrier
term and increases the gate fidelity. First,  
we describe the method itself in the subsection (\ref{subsec:pulse_shaping}).
Then,  we demonstrate its performance on a particular example
of a 5-ion chain in (\ref{subsec:pulse_shaping_example}). 
We show that our method offers a signifiant fidelity gain in 
comparison  with the pulse obtained from the linear 
Eqs.~\eqref{eq:gate_condition_alpha_nocar}, 
\eqref{eq:gate_condition_chi_nocar}, which are used
in the most of the pulse shaping approaches.
After that, in the subsection~\ref{subsec:mu_t_dep}, 
we study the applicability of our method for other 
gate durations, chain lengths, and values of bichromatic detuning.

\subsection{Pulse shaping scheme} \label{subsec:pulse_shaping}

Assuming that $\Omega(t)$ satisfies the smoothness conditions from 
Section~\ref{sec:evo_op_w_carrier}, the $\cos{2\Phi(t)}$ term in Eq.~\eqref{eq:f} 
can be 
averaged over fast oscillations on the timescale $\mu^{-1}$, as shown in 
Appendix~\ref{appendix:eff_omega}. As a result, we get an approximate 
expression for $f_{im}$ in Eq.~\eqref{eq:f}:
\begin{equation}
  f_{im}\approx \eta_{im}e^{i\omega_m t}\Omega_\mathrm{eff}(t)\cos{(\mu t + \psi)},
  \label{eq:f_eff}
\end{equation}
where
\begin{equation}
\Omega_\mathrm{eff}(t) = \mathcal{S}(\Omega(t)) = \Omega(t)\left(J_0\left(\frac{2\Omega(t)}{\mu}\right)  + J_2\left(\frac{2\Omega(t)}{\mu}\right)\right),
  \label{eq:eff_omega}
\end{equation}
and $J_k$ are Bessel functions. So, the $\cos{2\Phi(t)}$ term in Eqs.\eqref{eq:alpha_w_carrier}, 
\eqref{eq:chi_w_carrier} causing the nonlinearity of 
Eqs.~\eqref{eq:gate_conditions} disappears. Thus, $\alpha_{im}$ and $\chi_{ij}$ can be calculated 
with Eqs.~\eqref{eq:alpha_no_carrier}, \eqref{eq:chi_no_carrier} as there were no carrier term, but with
$\Omega(t)$ replaced by $\Oeff(t)$.

The transformation $\mathcal{S}(\Omega)$ given by Eq.~\eqref{eq:eff_omega} can be thought 
as a nonlinear squeezing transformation of $\Omega(t)$.
In Fig.~\ref{fig:omega_eff}, we depict the functional dependence of $\mathcal{S}(\Omega)$ on $\Omega$.
When  $\Omega \ll \mu$, $\mathcal{S}(\Omega) \approx \Omega$. At larger $\Omega \sim \mu$, 
$\mathcal{S}(\Omega)$ grows slower than $\Omega$ until it reaches 
its absolute maximum $\max(\mathcal{S}) = C\mu$, where $C\approx 0.581865$. 
Thus, the presence of the carrier term effectively reduces the field 
amplitude acting on an ion qubit. 

\begin{figure*}
  \centering
  \includegraphics[width=\linewidth]{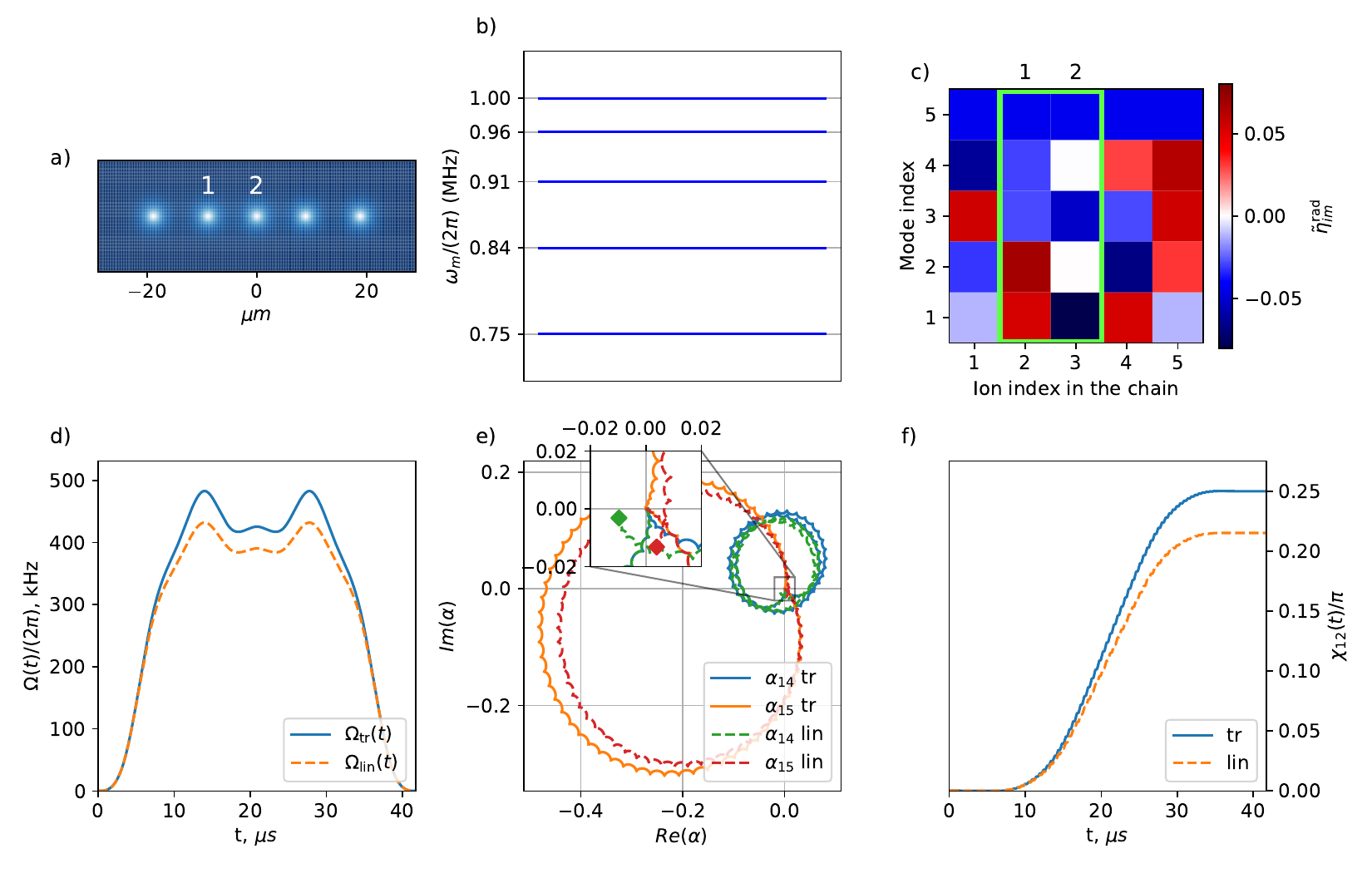}
  \caption{(a) Calculated ion equilibrium positions,  (b) radial normal mode frequencies of the chain, and
    (c) radial Lamb-Dicke parameter matrix for a chain of $\Ca$ ions.
    The indices $1,2$ in (a) and (c) indicate the ions illuminated by
    laser field. The green rectangle in (c) indicates two rows of the 
    Lamb-Dicke parameter matrix which correspond to the illuminated ions.
    (d) Non-optimized and optimized laser pulses applied to ions 1 and 2 indicated 
    in (a).
    (e) Phase trajectories of the ion 1 for modes 4 (stretch) 
    and 5 (COM) for both pulse shapes.  (f) Spin-spin 
    entanglement phase $\chi_{12}(t)$ between the illuminated 
  ions for both pulse shapes.}
  \label{fig:modes_and_pulse_5ions}
\end{figure*}

\begin{table}
  \centering

  \begin{tabular}{c|c|c}
      \toprule
      & $\Omega_\mathrm{tr}(t) $ & $\Omega_\mathrm{lin}(t) $\\
      \midrule
      $1-F_0$ &$1.4\times10^{-6}$ & $1.2\times10^{-2}$\\
      Num. (LD)   &$1.7\times10^{-6}$ & $1.2\times10^{-2}$\\
      Num. (full) &$5.7\times10^{-5}$ & $1.4\times10^{-2}$\\
  
  \bottomrule 
  \end{tabular}

  \caption{$\Rxx(\pi/4)$ gate fidelity for the pulses $\Omega_\mathrm{lin}$ and $\Omega_\mathrm{tr}$ with the initial state
    $|11\rangle_z$.  The values in the first row are obtained from the analytical expression 
    \eqref{eq:inf0_z}. The values in the second and the third row
  are obtained from TDSE solution with the Hamiltonians \eqref{eq:ld_ham} and \eqref{eq:full_ham}.}
  \label{tab:ms_results_z}
\end{table}

\begin{table*}

  \begin{tabular}{c|c|c|c|c|c|c|c}
      \toprule
      \multicolumn{2}{c|}{} & \multicolumn{3}{|c|}{$\Omega_\mathrm{tr}$} & \multicolumn{3}{|c}{$\Omega_\mathrm{lin}$}\\
      \midrule
      \multicolumn{2}{c|}{} & Eq.~\eqref{eq:inf_x} & Num. (LD) & Num. (full) & Eq.~\eqref{eq:inf_x} & Num. (LD) & Num. (full)\\
      \midrule
      \multirow{3}{*}{$|1,1\rangle_x$}  & $P_\mathrm{ph}^s$ &$1.9\times10^{-6}$ & $8.5\times10^{-7}$ & $3.4\times10^{-6}$ & $8.8\times10^{-4}$ & $9.0\times10^{-4}$ & $9.7\times10^{-4}$\\
      \cline{2-8}
                                        & $P_\mathrm{flip}^s$   &$9.3\times10^{-7}$ & $8.7\times10^{-7}$ & $6.6\times10^{-6}$ & $3.6\times10^{-7}$ & $3.3\times10^{-7}$ & $3.6\times10^{-6}$\\
      \cline{2-8}
                                        &  $1 - F_\mathrm{tot}$ &$2.8\times10^{-6}$ & $1.8\times10^{-6}$ & $1.0\times10^{-5}$ & $8.8\times10^{-4}$ & $9.0\times10^{-4}$ & $9.7\times10^{-4}$\\
      \cline{1-8}
      \multirow{3}{*}{$|1,-1\rangle_x$}    & $P_\mathrm{ph}^s$ &$6.3\times10^{-7}$ & $4.6\times10^{-7}$ & $1.1\times10^{-6}$ & $1.1\times10^{-4}$ & $1.1\times10^{-4}$ & $1.0\times10^{-4}$\\
      \cline{2-8}                                               
                                       & $P_\mathrm{flip}^s$   &$5.7\times10^{-7}$ & $5.3\times10^{-7}$ & $3.5\times10^{-6}$ & $2.5\times10^{-7}$ & $2.2\times10^{-7}$ & $2.2\times10^{-6}$\\
      \cline{2-8}                                               
                                       &  $1 - F_\mathrm{tot}$ &$1.2\times10^{-6}$ & $1.2\times10^{-6}$ & $4.7\times10^{-6}$ & $1.1\times10^{-4}$ & $1.1\times10^{-4}$ & $1.1\times10^{-4}$\\
  
  \bottomrule 
  \end{tabular}

  \caption{Contributions into $\Rxx(\pi/4)$ gate fidelity for the pulses $\Omega_\mathrm{lin}$ and $\Omega_\mathrm{tr}$ 
    with the initial states $|1,\pm1\rangle_x$. Each data row of the table corresponds to a contribution to the infidelity 
    with the initial state given in the leftmost column. Each data column of the table corresponds to a calculation method 
    (analytical or numerical with one of the considered Hamiltonians) for the pulse shape given in the top row.}
  \label{tab:detailed_ms_results_x}
\end{table*}

Using the transformation \eqref{eq:eff_omega}, we can find approximate 
solutions $\Omega_\mathrm{tr}(t)$ of the nonlinear 
equations \eqref{eq:gate_conditions} in two simple steps:
\begin{enumerate}
  \item Find $\Omega_\mathrm{lin}(t)$ satisfying 
    Eqs.~\eqref{eq:gate_condition_alpha_nocar}, \eqref{eq:gate_condition_chi_nocar} and smoothness conditions, 
    where $\alpha^0_{im}$ and $\chi^0_{ij}$ are defined by 
    Eqs.~\eqref{eq:alpha_no_carrier}, \eqref{eq:chi_no_carrier}, \eqref{eq:f0_0}. This pulse would implement the $\Rxx(\phi)$ gate
if the carrier term was not present. 
  \item     Find the solution by applying the inverse transformation 
    $\mathcal{S}^{-1}$:
    $\Omega_\mathrm{tr}(t) = \mathcal{S}^{-1}(\Omega_\mathrm{lin}(t))$.
    This accounts for the modified expressions for the spin-dependent forces.
\end{enumerate}

The above method was derived with the assumption that the averaging procedure of the Appendix~\ref{appendix:eff_omega} is applicable. 
However, this is not obvious for the short gate durations. 
Because of that, in the two following subsections we verify our method without initial assumptions on $\Omega(t)$ and analyse 
its applicability for different gate parameters. 
In the analysis, we use the analytical expressions for fidelity of the Section~\ref{sec:carrier_infidelity} and the numerical solution of the time-dependent Schrödinger equation (TDSE). 
We demonstrate that the second step of our procedure (transformation $\mathcal{S}^{-1}$) leads to a considerable fidelity increase. 
Although we consider a particular family of pulse shapes $\Omega_\mathrm{lin}$ represented by cubic splines, we expect that the effect of $\mathcal{S}^{-1}$ holds for a wider class of pulse shapes.

\subsection{5-ion chain example}
\label{subsec:pulse_shaping_example}

Let us illustrate the presented pulse shaping scheme on a particular example. 
We consider a chain of $n_\mathrm{ions}=5$ $\Ca$ ions in a harmonic pseudopotential
with the radial frequency of $1$ MHz and the axial frequency 
of $264.8\mathrm{kHz}$. The frequency of the lowest radial mode at these parameters 
is $0.75\mathrm{MHz}$.
We calculate the ion equilibrium positions, phonon normal modes and frequencies and 
the Lamb-Dicke parameters of the chain using standard 
methods (see Fig.\ref{fig:modes_and_pulse_5ions}a-c and
Appendix~\ref{appendix:normal_modes}) \cite{James1998}.
Then, we find the pulses implementing the $\Rxx(\phi=\pi/4)$ gate between the second and the 
third ions of the chain for the gate duration $41.74$ $\mu$s, the detuning $\mu = 2\pi\times 1.034\,\mathrm{MHz}$, 
and the motional phase $\psi=0$.

We search for $\Omega_\mathrm{lin}(t)$ as a smooth piecewise-cubic polynomial (qubic spline)
with $2n_\mathrm{ions}+2 = 12$ segments equally spaced in the interval time $(t_0, t_f)$ with $t_0 = 0$ and $t_f = 41.74$ $\mu$s.
Additionally, we require that the values and the derivatives 
vanish at the beginning and at the end of the gate.
Under these requirements, Eqs.~\eqref{eq:gate_condition_alpha_nocar}, 
\eqref{eq:gate_condition_chi_nocar} have a unique solution 
(see Appendix~\ref{appendix:spline_pulse_shaping}). To find
$\Omega_\mathrm{tr}(t)$, we apply the inverse 
transformation \eqref{eq:eff_omega}. For the considered 5-ion chain, $\Omega_\mathrm{lin}(t)$ and 
$\Omega_\mathrm{tr}(t)$ are shown in Fig.~\ref{fig:modes_and_pulse_5ions}d.

Then, we compare the system dynamics 
for $\Omega_\mathrm{lin}(t)$ and $\Omega_\mathrm{tr}(t)$.
Using Eqs.~\eqref{eq:alpha_w_carrier}, \eqref{eq:chi_w_carrier}, we find $\alpha_{im}(t, t_i)$ and $\chi_{ij}(t, t_i)$ 
entering the propagator \eqref{eq:propagator} for both pulses. For the center-of-mass (COM) and stretch modes of an ion 
crystal which are excited the most during the gate operation, the phase trajectories are shown in 
Fig.~\ref{fig:modes_and_pulse_5ions}e, and $\chi_{12}(t)$ is shown in Fig.~\ref{fig:modes_and_pulse_5ions}f. 
From the inset in Fig.~\ref{fig:modes_and_pulse_5ions}e, it is clear that the phonon mode trajectories are not perfectly 
closed for the pulse $\Omega_\mathrm{lin}(t)$ (the phonon mode amplitudes are of the order of $\sim 10^{-2}$ in the end of the gate). 
In contrast, for $\Omega_\mathrm{tr}(t)$ they are almost perfectly closed, and the deviation of the amplitudes from zero is 
invisible on the plot. 
Also, the deviation of $\chi_{12}$ at the end of the gate is significant for $\Omega_\mathrm{lin}(t)$, as opposed to $\Omega_\mathrm{lin}(t)$ (invisible on the plot).
Then, we use the abovementioned results to calculate the zero-order infidelity $F_0$ (Eq.~\eqref{eq:inf0_z}) 
for the initial states $|11\rangle_z$. The results are presented in Table~\ref{tab:ms_results_z}.
One can see that the theoretical infidelity for the 
pulse $\Omega_\mathrm{tr}$ is $\sim 10^{-6}$,
whereas the infidelity for the pulse $\Omega_\mathrm{lin}$ is of the order of $10^{-2}$.

We verify our analytical predictions by solving TDSE numerically for the full Hamiltonian \eqref{eq:full_ham} using QuTiP \cite{QuTiP2}. The calculated infidelities for both pulses are also in Table~\ref{tab:ms_results_z}. Qualitatively, the numerical results confirm our analytical prediction that the transformed pulse greatly reduces the gate error.  Quantitatively, we obtain good agreement for $\Omega_\mathrm{lin}$ and sligtly larger discrepancies for $\Omega_\mathrm{tr}$. We attribute the difference for
$\Omega_\mathrm{tr}$ to the influence of the higher-order terms in the Lamb-Dicke expansion, which are not accounted in our calculations but are present in the full Hamiltonian. To confirm this, we solve TDSE for the LD Hamiltonian \eqref{eq:ld_ham} where the higher-order terms of the expansion are neglected. The resulting infidelities, also presented in Table I, show much better agreement with the analytical predictions.

Similarly, we analyze the gate fidelity for the initial states $|1,\pm 1\rangle_x$. For these states, 
Eq.~\eqref{eq:inf_x} allows to separate the contributions of imperfectly closed phase trajectories and spin flips. These contributions can be identified from the numerical solution of TDSE: 
the probability of phonon mode excitation corresponds to the first term in Eq.~\eqref{eq:inf_x}, and the probability of the spin flip corresponds to the second term in Eq.~\eqref{eq:inf_x}.
In Table~\ref{tab:detailed_ms_results_x}, we present the results of analytical 
calculations and the TDSE solution both with Hamiltonians \eqref{eq:full_ham} and \eqref{eq:ld_ham}.

The results confirm that the first-order perturbation theory in $\hat{V}$ gives an accurate description of the dynamics of the LD Hamiltonian. We obtain good agreement between the analytical and numerical predictions for all cases except for the $P^s_\mathrm{ph}$ for the transformed pulse. Even in this case, the analytical formula correctly predicts the very small magnitude of phonon excitation probability $\sim 10^{-6}$.

In the simulation of the full Hamiltonian \eqref{eq:full_ham} with 
pulses $\Omega_\mathrm{lin}$, the considered contributions
into error also agree with the analytical predictions. 
However, for $\Omega_\mathrm{tr}$, both of the contributions 
are by order of magnitude larger than the corresponding analytical predicions. 
Still, the total error for $\Omega_\mathrm{tr}$ is  more than by order of 
magnitude lower than for $\Omega_\mathrm{lin}$.

From these results, we conclude that the dominant contributions into gate error for $\Omega_\mathrm{lin}$ are the error $\Delta\chi_{12}$ in the spin
coupling phase
and the imperfect closing of the phase trajectories as in Eq.~\eqref{eq:inf0_z}. Both of these contribuions can be almost canceled by transforming $\Omega_\mathrm{lin}$ into $\Omega_\mathrm{tr}$. 
The spin-flip contribution into error is negligible for both pulses. Thus, the pulse 
$\Omega_\mathrm{tr}$ allows the implementation of a fast high-fidelity $\Rxx(\pi/4)$ gate in a 5-ion chain.

\subsection{Gate time, detuning, and chain length dependencies}
\label{subsec:mu_t_dep}

\begin{figure*}[t]
  \centering
  \includegraphics[width=\linewidth]{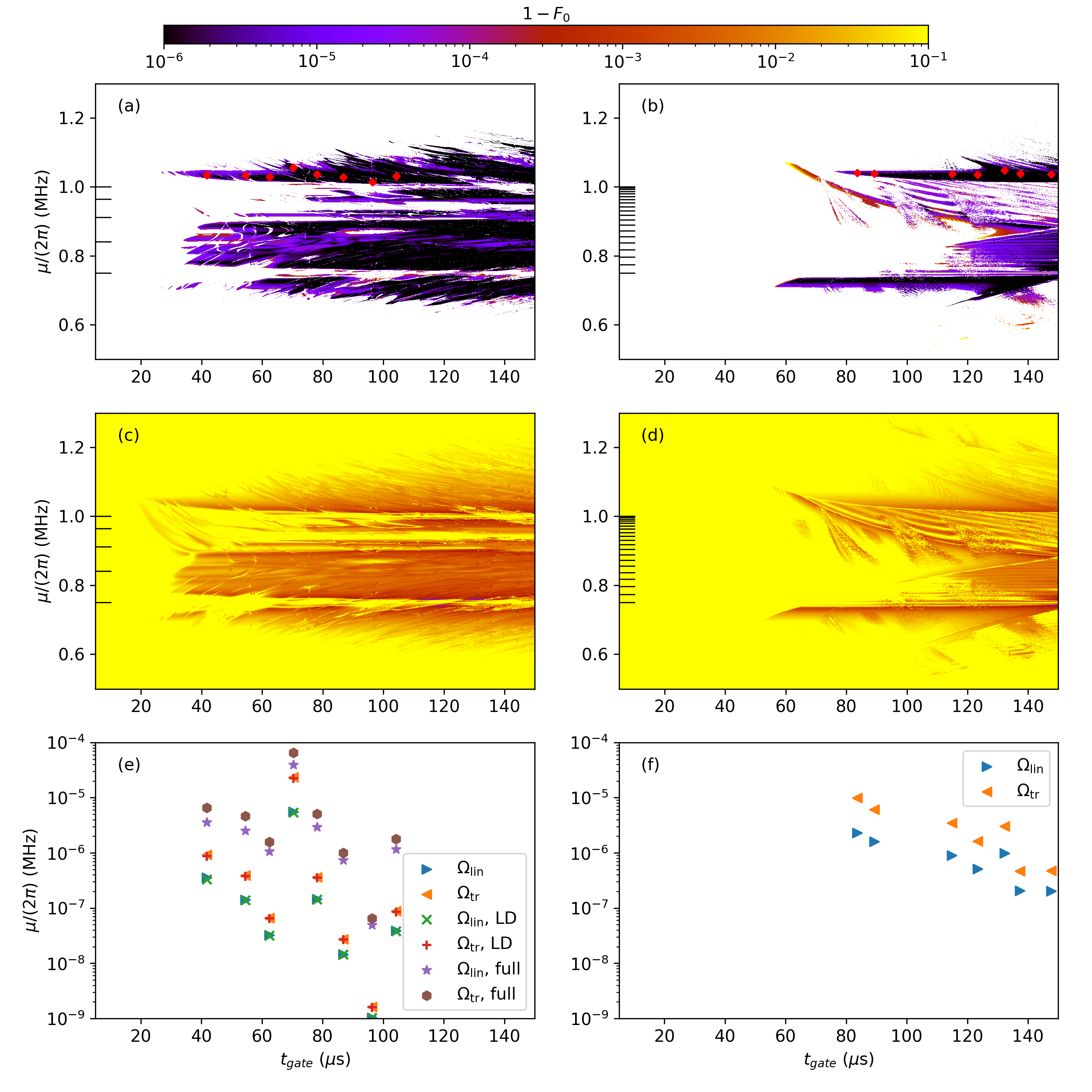}
  \caption{(a,b) Leading-order infidelity $1-F_0$ (indicated by color) for transformed pulses 
    $\Omega_\mathrm{tr}$ inside the area where 
    the inverse transformation $\mathcal{S}^{-1}$ exists for 5-ion (a) and 
    20-ion (b) chains. White color outside the allowed area indicates that $\mathcal{S}^{-1}$ is not defined.
    The red dots specify the values of $(t_\mathrm{gate}, \mu)$ for 
    which the simulations are performed.  (c, d) 
    Leading-order infidelity $1-F_0$ for non-transformed pulses 
    $\Omega_\mathrm{lin}$ for 5-ion (a) and 20-ion (b) chains. The black 
    horizontal lines in (a-d) indicate the positions of normal mode 
    frequencies.
    (e) Spin flip error for the initial state $|11\rangle_x$ of a 5-ion chain with
   $(t_\mathrm{gate}, \mu)$ labeled by dots in (a). It is calculated analytically and numerically 
   with the LD and full Hamiltonians for transformed and non-transformed
   pulses.
   (f) Analytical spin flip error for the initial state 
   $|11\rangle_x$ of a 5-ion chain with $(t_\mathrm{gate}, \mu)$ specified by dots in (b). 
    }
  \label{fig:allowed_areas}
\end{figure*}

\begin{figure}[h]
  \centering
  \includegraphics[width=\linewidth]{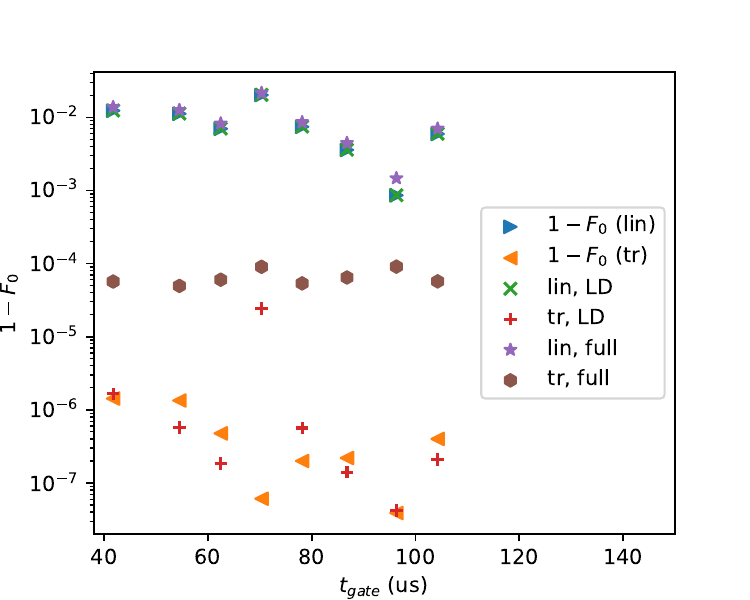}
  \caption{Leading-order infidelity $1-F_0$ compared with the infidelity calculated from numerical simulation for the initial state $|11\rangle_z$ in 5-ion chain 
    with $t_\mathrm{gate},\mu$ specified in Fig.~\ref{fig:allowed_areas}(a). The values of $1-F_0$ are compared with the simulation results for the LD Hamiltonian and full Hamiltonian both for transformed and non-transformed pulses.}
  \label{fig:z_states_inf}
\end{figure}

\begin{figure}[h]
  \centering
  \includegraphics[width=\linewidth]{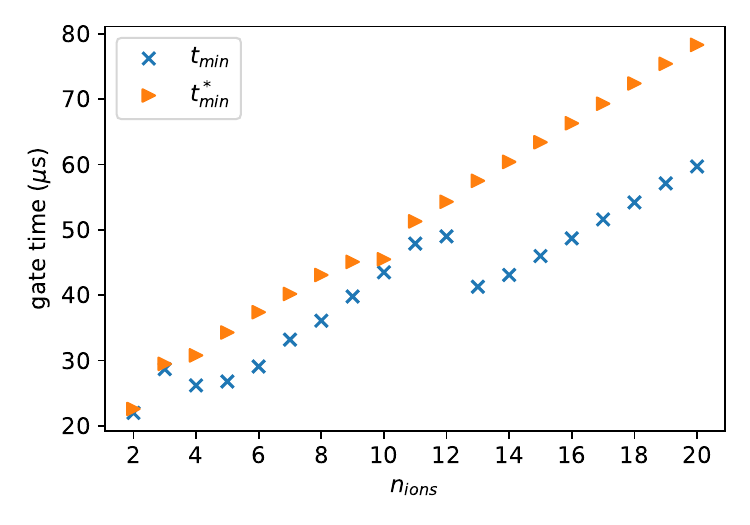}
  \caption{
    Minimal gate duration $t_\mathrm{min}$ 
  for which $\Oeff$ does not exceed $C\mu$ 
  and minimal gate duration $t_\mathrm{min}^*$ 
  ensuring the zero-order error equal to $10^{-5}$ for
  optimized pulses and $|\psi_0\rangle = |11\rangle_z$ 
  as functions of $n_\mathrm{ions}$. 
}
  \label{fig:n_ions_dep}
\end{figure}

Here we examine the applicability of the method for 
other gate durations, detunings, and chain lengths. Below, we find the parameter domains
for which the method of Section~\ref{sec:pulse_shaping} is applicable. For the 
5-ion chain considered previously, we peform analytical and numerical calculations 
for different parameter values and study the fidelity dependence on these parameters. 
For longer ion chains, we perform only analytical calculations because they require an exponentially 
growing amount of computational resources for numerical modelling.

A key point for the implementation of our scheme is the existence of the inverse transformation $\mathcal{S}^{-1}$. As shown in the subsection~\ref{subsec:pulse_shaping} (see Eq.~\eqref{eq:eff_omega} and Fig.~\ref{fig:omega_eff}), the effective field amplitude $\Oeff$ has an absolute maximum of $C\mu$: therefore, $\mathcal{S}^{-1}$ is defined only in the range $(-C\mu, C\mu)$. So, for the step 2 of the pulse shaping procedure in the subection~\ref{subsec:pulse_shaping}, it is necessary that the absolute value of $\Omega_\mathrm{lin}$ obtained from linear equations does not exceed 
$C\mu$. This requirement poses a restriction on the possible values of the system and the gate parameters.

The pulse shapes $\Omega_\mathrm{lin}$ obtained from linear equations 
Eq.~\eqref{eq:gate_condition_alpha_nocar}, \eqref{eq:gate_condition_chi_nocar}
depend on the number of ions, the values of normal mode frequencies, the Lamb-Dicke parameters, the detuning of the bichromatic beam, 
and the gate duration.
 The normal mode frequencies and the Lamb-Dicke 
parameters depend on the trap frequencies, ion masses and the wavevectors. However, these dependencies  are rather 
simple and do not contain fastly oscillating contributions. 
 The dependence on the motional phase $\psi$ is weak provided that the field amplitude is smoothly turned on and off. In contrast, the dependence on the gate time $t_\mathrm{gate}$ and $\mu$ is highly non-trivial because they enter the pulse shaping equations inside the oscillatory integrals \eqref{eq:A_matrix}.

Therefore, we mostly focus on the dependence on $t_\mathrm{gate}$ and $\mu$. We find the sets in the plane $(t_\mathrm{gate}, \mu)$ where $\mathcal{S}^{-1}$ exists 
 ($\Omega_\mathrm{lin}$ does not exceed  $C\mu$), which we call the allowed areas (see Fig.~\ref{fig:allowed_areas}). 
Due to complicated dependence of $\Omega_\mathrm{lin}$ on $t_\mathrm{gate}$ and $\mu$, 
the allowed areas have complicated shape in the $t_\mathrm{gate}$-$\mu$ plane.

First, we find the allowed area (see Fig.~\ref{fig:allowed_areas}a) for the 5-ion chain considered in the previous 
section. We calculate $\Omega_\mathrm{lin}(t)$ for a  ($1450\times 800$)
grid in $t_\mathrm{gate}$-$\mu$ plane for gate durations 
from $5$ to $150$ $\mu$s and detunings from $0.6$ to $1.2 \mathrm{MHz}$. 
Along 
the vertical axis, the area occupies the values of $\mu$ lying closely to the 
band of radial phonon frequencies. Along the horizontal axis, it spans the whole 
range of $t_\mathrm{gate}$ except the values below $\sim 30 \mu\mathrm{s}$. 
For each grid point inside the allowed area, we calculate the transformed pulse
$\Omega_\mathrm{tr}(t)$. Using Eq.~\eqref{eq:inf0_z}, we calculate the leading-order 
contribution $1-F_0$ to gate infidelity for the initial state
$|11\rangle_z$ for each pair $t_\mathrm{gate}, \mu$. The result is shown by color
inside the allowed area in Fig.~\ref{fig:allowed_areas}a. We find that infidelity is of the order of 
$10^{-5}$ nearly for all the grid points inside the area except the vicinity of the
boundaries. Larger infidelity values near the allowed area boundaries 
indicate that the slowly-varing approximation used to derive the Eq.~\eqref{eq:eff_omega} becomes less accurate in this case.

For comparison, we calculate $1-F_0$ 
for pulses $\Omega_\mathrm{lin}$ for the initial state $|11\rangle_z$ for the entire 
grid. The result is shown in Fig~\ref{fig:allowed_areas}(c). We find that the 
infidelity for $\Omega_\mathrm{lin}$ is of the order of $10^{-3}$-$10^{-2}$ inside the allowed area,
which is considerably larger than the infidelity for $\Omega_\mathrm{tr}$. Outside the allowed area,
the error reaches $\sim 10^{-1}$ and even more due to larger values of 
$\Omega_\mathrm{lin}$.

We repeat these calculations for a 20-ion chain (see the results in Fig.~\ref{fig:allowed_areas}(b,d). 
All laser and trap parameters
are taken the same as for the 5-ion case except the axial frequency and the number of segments in the pulse.
The number of segments in the pulse is enlarged to $2n_\mathrm{ions}+1 = 41$.
The axial frequency is set to  $78.7\,\mathrm{kHz}$, which results in radial phonon modes 
lying in the range $(0.75\mathrm{MHz}, 1\mathrm{MHz})$ similarly to the 5-ion case.
In Fig.~\ref{fig:allowed_areas}(b,d), we show the results 
for a 20-ion chain. Analogously to the 5-ion case, we find the allowed area and 
calculate $1-F_0$ for $\Omega_\mathrm{lin}$ and $\Omega_\mathrm{tr}$ in the 
$t_\mathrm{gate},\mu$ plane. 
The 20-ion allowed area (see Fig.~\ref{fig:allowed_areas}(b)) has a shape similar to the 5-ion case,
however, it is shifted to larger gate durations: the minimal time for points inside 
the area is $\sim 60\mu\mathrm{s}$.
Analogously with the 5-ion case, 
we find that $1-F_0$ is of the order $10^{-5}$ for $\Omega_\mathrm{tr}$ (see Fig.~\ref{fig:allowed_areas}(b)) and of the order of 
$10^{-2}$ for $\Omega_\mathrm{lin}$ (see Fig.~\ref{fig:allowed_areas}(d)). 

For other chain lengths from 2 to 20, we get similar results. 
In Fig.~\ref{fig:n_ions_dep}, we show the dependence of the minimal time  $t_\mathrm{min}$ 
inside the allowed area on the number of ions. For each $n_\mathrm{ions}$, we find the 
minimal gate duration $t_\mathrm{min}^*$ for which $1-F_0$ reaches $10^{-5}$. Both of these
durations grow monotonically with the increasing number of ions, with 
$t_\mathrm{min}^*$ exceeding $t_\mathrm{min}$ no more than by $\sim 20 \mu\mathrm{s}$.

Also, we calculate analytically the spin flip probability for a 
selected set of points in $t_\mathrm{gate},\mu$ plane within the allowed area
for 5-ion and 20-ion chains 
for the initial state $|11\rangle_x$. The points are shown as dots
in Fig.~\ref{fig:allowed_areas}(a) (5 ions) and 
Fig.~\ref{fig:allowed_areas}(b) (20 ions). The values of the spin flip
probability are shown in Fig.~\ref{fig:allowed_areas}(e) and 
Fig.~\ref{fig:allowed_areas}(f). For all selected points, the spin flip 
probability for a 5-ion chain does not exceed $10^{-4}$, and it is even 
smaller for a 20-ion chain.

In addition, we perform numerical simulations for a selected set of points 
shown by dots in Fig.~\ref{fig:allowed_areas}(a) for a 5-ion chain. For each 
pair $(t_\mathrm{gate}, \mu)$ indicated by a dot, we model gate dynamics by solving 
TDSE with the pulses $\Omega_\mathrm{lin}(t)$ and $\Omega_\mathrm{tr}(t)$ 
for the initial states $|11\rangle_z$ and $|11\rangle_x$. 

For the initial states $|11\rangle_x$, we calcucate the spin flip error 
as in Section~\ref{sec:pulse_shaping}. We compared the numerical simulations of the  
full Hamiltonian, of LD Hamiltonian, and the perturbation theory.  The results of these 
calculations confirm the findings of the Section~\ref{sec:pulse_shaping}: 
perturbation theory gives an accurate result for the  
LD Hamiltonian \eqref{eq:ld_ham}, but the spin flip probability $P_\mathrm{flip}^s$ 
is by order of magnitude higher for the full Hamiltonian \eqref{eq:full_ham}, which is of the order $10^{-5}$. 

For the initial states $|11\rangle_z$, we perform calculations similar to
that of Table~\ref{tab:ms_results_z}. We compare the analytical zero-order infidelity 
given by Eq.~\eqref{eq:inf0_z} with the infidelity obtained from TDSE. The results 
are presented in Fig.~\ref{fig:z_states_inf}. For all points, the infidelity for 
the pulses $\Omega_\mathrm{tr}$  is considerably lower than for 
$\Omega_\mathrm{lin}$. By comparing the numerical indidelities
for the full Hamiltonian and the LD Hamiltonian, 
we conclude that the Eq.~\eqref{eq:inf0_z}
gives a leading-order contribution to the error for $\Omega_\mathrm{lin}$.
In contrast, he dominant contribution to the error for $\Omega_\mathrm{tr}$
originates from the higher orders of the Lamb-Dicke expansion.

These results show that the conclusions of the Section~\ref{sec:pulse_shaping} 
persist for a wide range of these parameters providing that 
$t_\mathrm{gate}$ and $\mu$ are within the allowed area.
Also, the calculations prove that the zero-order error can be 
reliably used to estimate the fidelity gain given by our scheme.

For gate durations of tens of microseconds, our scheme allows to reduce 
the gate error from $\sim 10^{-2}$ for non-optimized pulses to $\sim 10^{-5}$. 
Thus, our scheme facilitates the implementation of fast high-fidelity 
MS gates in chains with tens of ions, which is crucial for speeding up trapped-ion
quantum computation.

\section{Conclusions}
We propose an amplitude-pulse-shaping method to shorten the M{\o}lmer-S{\o}rensen entangling gates in linear chains of 
trapped-ion qubits. Our method allows to considerably reduce the error originating from the carrier transition, which is an intrinsic 
feature of the ions interaction with bichromatic external field. 
We show that a certain nonlinear transformation of the laser pulse 
allows to compensate the modification of the spin-dependent forces by the carrier transition almost entirely, thus enabling a fast high-fidelity gate. 
For short ion chains, we show analytically and numerically that our method allows to reduce gate infidelity 2-3 orders of magnitude for 
gate durations of tens of microseconds
with the resulting fidelity below $\sim 10^{-4}$. According to our analytical calculations, these conclusions persist for chains at least up to 20 ions.
 
\begin{acknowledgments}
This work was supported by Rosatom in the framework of
the Roadmap for Quantum computing (Contract No. 868-1.3-15/15-2021).
\end{acknowledgments}

\section*{Data availability}
The data that support the findings of this article are openly available \cite{FastMolmerSorensenZenodo}.

\appendix

\section{The MS gate fidelity with account of first-order spin-flip correction}
\label{appendix:fidelity_expr}
In this Appendix, we derive expressions for the M{\o}lmer-S{\o}rensen gate fidelity with account to three contributions: incompletely closed phonon mode phase 
trajectories, error in spin coupling phase, and spin-flip contribution.  We consider fidelity in the full ion-phonon Hilbert space assuming that 
all phonon modes are initially cooled to the ground state. So, fidelity $F_\mathrm{tot}$ is defined by Eq.~\eqref{eq:fidelity_def}
as squared overlap between the target state and the resulting state.
To find $F$ using perturbation theory, it is convenient to denote the combination $\Rxx^\dagger(\phi) U$ as 
\begin{equation}
  \Rxx^\dagger(\phi) \hat{U} = 1 - i\hat{T}.
\end{equation}
Using the unitarity of $\Rxx^\dagger(\phi) \hat{U}$, we can rewrite fidelity in the form more suitable for perturbative calculation:
\begin{equation}
  1 - F_\mathrm{tot} = \langle\psi_0 |\hat{T}^\dagger \hat{T} |\psi_0\rangle - \langle\psi_0 |\hat{T} |\psi_0\rangle^2.
  \label{eq:fidelity_from_T}
\end{equation}
Then, we use the MS gate propagator $\hat{U}$ with the first-order correction from the spin-flip perturbation \eqref{eq:pert_first_ord_U}.
Let us give a small-error expansion of $\hat{T}$. We assume that 
$\alpha_{im}$ and the deviation of $\chi_{ij}$ from the target values are small, so the MS propagator can be decomposed as 
\begin{equation}
  \Rxx(\phi)^\dagger \hat{U}_0 = \mathds{1} - i\delta \hat{U}_\MS,
  \label{eq:delta_u_ms_def}
\end{equation}
where 
\begin{equation}
  \delta \hat{U}_\MS \approx \frac{1}{2}\sum_{i,j}\chi_{ij}\sigma_x^i\sigma_x^j + i\sum_{im} \sigma_x^i(\alpha_{im}^*\ha_m - \alpha_{im}\ha^\dagger_m).
  \label{eq:delta_u_ms_expr}
\end{equation}
After that, $\hat{T}$ can be approximated as
\begin{equation}
  \hat{T} \approx \delta \Ums + \hat{T}_1, 
  \label{eq:T_w_carrier}
\end{equation}
where we define $\hat{T}_1$ by Eq.\eqref{eq:pert_first_ord_T}, and 
neglect the term $\delta U_\MS \hat{T}_{1}$. We find convenient to separate the contributions of $\delta\Ums$ and $\hat{T}_1$ into gate infidelity.
For that, we define two auxiliary fidelities. The first one is the fidelity $F_0$ of the state obtained by the action of the zero-order evolution operator \eqref{eq:propagator} defined in Section~\ref{sec:carrier_infidelity}, 
Eq.~\eqref{eq:F0_def}.
It contains only the contributions of the imperfectly closed phase trajectories and the error in spin coupling phase. The second one is the fidelity between the states
$\hat{U}_0|\psi_0\rangle$ and $\hat{U}|\psi_0\rangle$:
\begin{equation}
  F_c = |\langle \psi_0 | \hat{U}_0^\dagger \hat{U} |\psi_0\rangle|^2
  \label{eq:f_c}
\end{equation}
It characterizes the deviation of the full evolution operator $\hat{U}$ from the M{\o}lmer-S{\o}rensen propagator \eqref{eq:propagator} and is determined by $\hat{T}_1$. Similarly with
\eqref{eq:fidelity_from_T}, we can express $F_0$ and $F_c$ as 
\begin{equation}
  1 - F_0 = \langle\psi_0| \delta \Ums^\dagger \delta\Ums|\psi_0\rangle - |\langle\psi_0 |\delta\Ums| \psi_0\rangle|^2.
  \label{eq:inf_ms}
\end{equation}
and 
\begin{equation}
  1 - F_c = \langle\psi_0|\hat{T}_1^\dagger \hat{T}_1|\psi_0\rangle -
  |\langle\psi_0|\hat{T}_1|\psi_0\rangle|^2.
  \label{eq:inf_c}
\end{equation}
The auxiliary fidelities \eqref{eq:F0_def}, \eqref{eq:f_c} are useful for analysing the fidelity $F_\mathrm{tot}$ (Eq.~\eqref{eq:fidelity_def}). First, for any initial state, $F_\mathrm{tot}$ can be bounded using the triangle inequality for fidelities \cite{NielsenChuang2010}:
\begin{equation}
  F_\mathrm{tot} > \cos{(\arccos{F_0} + \arccos{F_c})}. 
  \label{}
\end{equation}
Second, let us prove that the infidelity \eqref{eq:fidelity_from_T} averaged over initial states is simply a sum of \eqref{eq:inf_ms} and \eqref{eq:inf_c}. 
Indeed, the average of \eqref{eq:fidelity_from_T} can be calculated as
\begin{multline}
  1 - \avg{F_\mathrm{tot}} = \frac{1}{2^{N}}\left[ \Tr_q \langle 0_{ph} | \hat{T}^\dagger \hat{T} | 0_{ph}\rangle \right. \\
  - \left. \Tr_q  \langle 0_{ph} | \hat{T}^\dagger | 0_{ph}\rangle \langle 0_{ph} |\hat{T} | 0_{ph}\rangle\right],
  \label{eq:avg_inf}
\end{multline}
where $\Tr_q$ denotes the trace over qubit space. Then, we substitute 
the Eq.~\eqref{eq:T_w_carrier} into Eq.~\eqref{eq:avg_inf}. 
As $\delta\Ums$ commutes with 
all $\sigma_x^i$, and $\hat{T}_1$ contains only terms flipping the pseudospin
direction along the $x$ axis, all cross-products of $\delta \Ums$ and
$\hat{T}_1$ vanish. Therefore, we get
\begin{equation}
  1 - \avg{F_\mathrm{tot}} = (1 - \avg{F_0}) + (1 - \avg{F_c}),
\end{equation}
Using Eq.~\eqref{eq:avg_inf}, we get the following 
expression for $\avg{F_0}$:
\begin{equation}
  1 - \avg{F_0} = \sum_{i=1,2} |\alpha_{im}|^2 + \frac{4}{5}\delta\chi_{12}^2.
\end{equation}
The contribution $F_c$ averaged over initial states
can be bounded from above as an average over all qubit basis states 
in the $x$-basis, which simply follows
from Eq.~\eqref{eq:avg_inf}:
\begin{equation}
  1 - F_c < \frac{1}{4} \Tr_{q} \hat{T}_1^\dagger \hat{T}_1 = 
  \frac{1}{4} \sum_s \langle s, 0_\mathrm{ph}|\hat{T}_1^\dagger \hat{T}_1 |s,0_\mathrm{ph}\rangle.
\end{equation}
A closed-form representation of the 
term $\langle s, 0_\ph|\hat{T}_1^\dagger \hat{T}_1 |s,0_\ph\rangle$,
which is the probability of a spin flip during the MS gate operation
for the initial state $|s,0_\ph\rangle$, is given in Appendix~\ref{appendix:spin_flip}.

Now let us give expressions for some particular initial states. First of all,
let us take the initial state of the form 
$|s\rangle\otimes|0_\mathrm{ph}\rangle$. Similarly with the average 
fidelity, the action of $\delta \Ums$ and $\hat{T}_1$ in the 
qubit space implies that the infidelity is a sum of contributions
from these operators. We get
\begin{equation}
  1 - F_0 = \sum_m \left|\sum_{i} \alpha_{im} s_i\right|^2,
\end{equation}
\begin{equation}
  1 - F_c = 
  \langle s, 0_\mathrm{ph}| T_1^\dagger T_1 |s, 0_\mathrm{ph}\rangle,
\end{equation}
\begin{equation}
  1 - F_\mathrm{tot} = \sum_m \left|\sum_{i} \alpha_{im} s_i\right|^2 + 
  \langle s, 0_\mathrm{ph}| T_1^\dagger T_1 |s, 0_\mathrm{ph}\rangle.
\end{equation}
Here $\delta\chi_{ij}$ does not contribute into error as it is responsible 
only for the phase between different qubit basis vectors in $x$-basis.

Finally, let us give expressions for $F_0$ for the initial state
$|\psi_0\rangle = |s\rangle_z\otimes|0_\ph\rangle$ (a spin string in $z$-basis) and for 
an arbitrary superposition of qubit states in $x$ basis:
\begin{equation}
  |\psi_0\rangle = \sum c_s|s\rangle\otimes|0_\mathrm{ph}\rangle.
  \label{eq:arb_init_state}
\end{equation}
For $|\psi_0\rangle = |s\rangle_z\otimes|0_\ph\rangle$, 
\begin{equation}
  1 - F_0 = \sum_{i=1,2} |\alpha_{im}|^2 + \delta\chi_{12}^2.
  \label{eq:ms_inf_z}
\end{equation}
For a superposition with arbitrary coefficients $c_s$,
\begin{multline}
  1 - F_0 = \sum_s |c_s|^2\left|\alpha^T s\right|^2 + 
  \frac{1}{4}\sum_s |c_s|^2 (s^T \delta\chi s)^2 \\- \frac{1}{4}\left|\sum_s |c_s|^2 (s^T \delta\chi s)\right|^2.
  \label{eq:ms_inf_arb_init_state}
\end{multline}
The expressions for $F_0$ for the latter states can be obtained in the 
same way. In general, they contain the interference terms coming 
from products of $\delta \Ums$ and $\hat{T}_1$.

\section{Spin flip probability}
\label{appendix:spin_flip}
In this Appendix, we derive a closed-form representation of the 
probability of a spin flip during the MS gate operation for the initial state $|s,0_\ph\rangle$,
which is given by the matrix element
$\langle s, 0_\mathrm{ph}|\hat{T}_1^\dagger \hat{T}_1 |s,0_\mathrm{ph}\rangle$,
We represent it as as a two-dimensional
integral. From Eq.~\eqref{eq:pert_first_ord_T}, we get
\begin{equation}
  \hat{T}_1^\dagger \hat{T}_1 = \int dt' dt'' \hat{U}_0^\dagger(t', t_0) \hat{V}(t') 
  \hat{U}_0(t',t'') \hat{V}(t'') \hat{U}_0(t'', t_0).
  \label{eq:t1t1}
\end{equation}
The propagator $\hat{U}_0$ given by Eq.~\eqref{eq:propagator} is diagonal in qubit space
and contains displacement operators in phonon space and can be represented as
\begin{equation}
  \hat{U}_0(t_2, t_1) = \sum e^{-i\chi_s(t_2, t_1)} D(\alpha^T(t_2, t_1) s) |s\rangle\langle s|,
  \label{}
\end{equation}
where $\alpha(t_2, t_1)$ is the matrix $\alpha_{im}(t_2, t_1)$, and 
\begin{equation}
  \chi_s(t_2, t_1) = \frac{1}{2}\sum_{ij} \chi_{ij}(t_2, t_1)s_i s_j.
\end{equation}
The perturbation $\hat{V}$ can be written as
\begin{equation}
  \hat{V}(t) = \sum \hat{V}_{m}^\beta(t) A_m^\beta, \quad \beta = 1,2,
\end{equation}
with
\begin{equation}
  \begin{gathered}
    \hat{A}^{1}_m = \hat{a}_m,\\
    \hat{A}^{2}_m = \hat{a}_m^\dagger,
  \end{gathered}
\end{equation}
\begin{equation}
  V_{m}^\beta(t) = \sum_i V_{im}^\beta(t) \sigma_z^i,
\end{equation}
\begin{equation}
  V_{im}^{1,2}(t) = \eta_{im}\Omega(t)\cos{(\mu t + \psi)} \sin{2\Phi(t)} e^{\mp i \omega_m t}.
\end{equation}
By substituting these expressions into Eq.~\eqref{eq:t1t1} and taking the required matrix element, one gets
\begin{widetext}
  \begin{multline}
    \langle s, 0_\ph|\hat{T}_1^\dagger \hat{T}_1 |s,0_\ph\rangle = \sum_{i,m_1,m_2,\beta_1,\beta_2,s'}\int dt_1 dt_2  
    (V_{m_1}^{\beta_1})_{ss'} (t_1)(V_{m_2}^{\beta_2})_{ss'} (t_2)
    e^{i(\chi_s(t_1,t_0) - \chi_{s'}(t_1, t_2) + \chi_s(t_2, t_0))}\\
    \langle 0_\ph| D^\dagger(\alpha^T(t_1, t_0)s)\hat{A}_{m_1}^{\beta_1}D(\alpha^T(t_1, t_2)s')\hat{A}_{m_2}^{\beta_2}
    D(\alpha^T(t_2, t_0)s)|0_\ph\rangle.
    \label{eq:spin_flip_error_explicit}
  \end{multline}
\end{widetext}
The matrix elements of the products of displacement operators and creation and annihilation operators can be calculated 
using the identites for displacement operators (see, for example, 
\cite{Scully1997}). Below we give the expressions for the matrix elements for all combinations of $\beta$, which corresponds
to all combinations of creation and annihilation operators:
\begin{widetext}
  \begin{equation}
    \begin{aligned}
      \langle 0_\ph| D^\dagger(\vec\alpha)\hat{a}_{m_1}D(\vec\alpha_2)\hat{a}_{m_2}
      D(\vec\alpha_3)|0_\ph\rangle &= 
      \langle 0_\ph| D^\dagger(\vec\alpha_1)D(\vec\alpha_2) D(\vec\alpha_3)|0_\ph\rangle
      ((\alpha_2)_{m_1} + (\alpha_3)_{m_1})(\alpha_3)_{m_2},\\
      \langle 0_\ph| D^\dagger(\vec\alpha_1)\hat{a}_{m_1}D(\vec\alpha_2)\hat{a}_{m_2}^\dagger
      D(\vec\alpha_3)|0_\ph\rangle &= 
      \langle 0_\ph| D^\dagger(\vec\alpha_1)D(\vec\alpha_2) D(\vec\alpha_3)|0_\ph\rangle
      (\alpha_1^*)_{m_1}(\alpha_3)_{m_3},\\
      \langle 0_\ph| D^\dagger(\vec\alpha_1)\hat{a}_{m_1}^\dagger D(\vec\alpha_2)\hat{a}_{m_2}
      D(\vec\alpha_3)|0_\ph\rangle &= 
        \langle 0_\ph| D^\dagger(\vec\alpha_1)D(\vec\alpha_2) D(\vec\alpha_3)|0_\ph\rangle
        (\delta_{m_1m_2} + ((\alpha_1)^*_{m_2} - (\alpha_2)^*_{m_2})((\alpha_2)_{m_1} + (\alpha_3)_{m_1})),\\
      \langle 0_\ph| D^\dagger(\vec\alpha_1)\hat{a}_{m_1}^\dagger D(\vec\alpha_2)\hat{a}_{m_2}^\dagger
      D(\vec\alpha_3)|0_\ph\rangle &= 
      \langle 0_\ph| D^\dagger(\vec\alpha_1)D(\vec\alpha_2) D(\vec\alpha_3)|0_\ph\rangle
      (\alpha_1)_{m_1}^*((\alpha_1)_{m_2}^* - (\alpha_2)^*_{m_2}),\\
      \langle 0_\ph| D^\dagger(\vec\alpha_1)D(\vec\alpha_2) D(\vec\alpha_3)|0_\ph\rangle &= \bigexp{-i\Im[(\vec\alpha_1\vec\alpha^*_2) + (\vec\alpha_1 \vec\alpha^*_3) - (\vec\alpha_2 \vec\alpha^*_3)]
      -\frac{1}{2}|\alpha_1 - \alpha_2 - \alpha_3|^2}.
    \end{aligned}
  \end{equation}
\end{widetext}

\section{Averaging the spin-dependent force over fast carrier oscillations}
\label{appendix:eff_omega}
The integrals for $\alpha_{im}$ and $\chi_{ij}$ entering the MS propagator 
(Eqs.~\eqref{eq:alpha_w_carrier} and \eqref{eq:chi_w_carrier}) contain the term $f_{im}$ (Eq.\eqref{eq:f}) with a 
fastly-oscillating multiplier $\cos{2\Phi(t)}$. Here we show how the fast oscillations can be approximately averaged
on the timescale $\sim \mu^{-1}$. 
For that, let us explicitly decompose $f_{im}$ as a product of 
slowly and fastly varying components:
\begin{equation}
  \label{eq:f_decomposition}
  f_{im}(t) = \underbrace{\frac{1}{2}\eta_{im}e^{i(\omega_m - \mu) t - i\psi}\Omega(t)}_\text{slow part}
  \underbrace{(1 + e^{2i(\mu t + \psi)})\cos{2\Phi(t)}}_\text{fast part}
\end{equation}
We assume that the slow part 
does not change significantly over several periods of carrier 
oscillations $\mu^{-1}$. Then, we can average the fast part assuming that 
$\Omega(t)$ is approximately constant. 
Also, due to the assumptions on $\Omega(t)$ made in 
Section~\ref{sec:pulse_shaping}, we can replace $\Phi(t)$ by the leading-order 
asymptotic contribution:
 \begin{equation}
   \Phi(t) = \int_{t_0}^{t} dt' \Omega(t') \cos{(\mu t' + \psi)} \approx 
   \frac{\Omega(t)}{\mu}\sin{(\mu t + \psi)}.
 \end{equation}
After that, the fast part can be averaged as follows:
\begin{multline}
  \avg{(1 + e^{2i(\mu t + \psi)})\cos{\left(\frac{2\Omega}{\mu}\sin{(\mu t + \psi})\right)}} = \\
  =J_0\left(\frac{2\Omega}{\mu}\right)  + J_2\left(\frac{2\Omega}{\mu}\right),
  \label{eq:bessel_avg}
\end{multline}
where $J_n$ are Bessel functions. 

By substituting \eqref{eq:bessel_avg} into \eqref{eq:f_decomposition}, we get 
\begin{multline}
  f_{im}(t) \approx \frac{1}{2}\eta_{im}e^{i(\omega_m - \mu) t - i\psi}\Omega(t)\times\\
  \underbrace{\left(J_0\left(\frac{2\Omega(t)}{\mu}\right)  + J_2\left(\frac{2\Omega(t)}{\mu}\right)\right)}_\text{averaged fast part}
\end{multline}
Due to smooth time dependence of $\Omega(t)$, we can replace the oscillating exponent with the cosine and obtain the Eq.~\eqref{eq:f_eff}.

\section{Normal modes of a linear trapped-ion crystal}
\label{appendix:normal_modes}
For a linear ion chain in a harmonic Paul trap potential, the 
equilibrium positions can be found from numerical minimization of the potential energy
\begin{equation}
  U = \sum_k\frac{m\omega_\mathrm{ax}^2 x_k^2}{2} + \sum_{k<l} \frac{e^2}{4\pi\epsilon_0|x_k-x_l|}.
  \label{}
\end{equation}
Then, normal modes and frequencies can be found from the second-order expansion of the potential near the equilibrium positions.
For the axial (radial) modes, the 
eigenfrequencies $\omega_\mathrm{ax}^{m}$ ($\omega_{\mathrm{rad}}^m$)
and the normal vectors
$b_{im}^\mathrm{ax}$ ($b_{im}^\mathrm{rad}$)
can be found from the following eigenvalue problems:
\begin{equation}
  \sum_l\left[M\omega_\mathrm{ax}^2\delta_{kl} + \frac{e^2}{2\pi\epsilon_0} \tilde G_{kl}\right]b^\mathrm{ax}_{lm} = 
  m(\omega_m^\mathrm{ax})^2 b_{km}^\mathrm{ax},
\end{equation}
\begin{equation}
  \sum_l\left[M\omega_\mathrm{rad}^2\delta_{kl} - \frac{e^2}{4\pi\epsilon_0} \tilde G_{kl}\right]b^\mathrm{rad}_{lm} = 
  M(\omega^\mathrm{rad}_m)^2 {b^\mathrm{rad}_{km}},
\end{equation}
where $\tilde{G}_{ij}$ is the Hessian matrix of the Coulomb potential:
\begin{equation}
  \tilde G_{kl} = \sum_{k'} \frac{1}{|x_k-x_{k'}|^3}\delta_{kl} -
\frac{1}{|x_k-x_l|^3}.
\end{equation}
The Lamb-Dicke parameters for the axial and radial
normal modes are defined as 
\begin{equation}
  \tilde\eta_{km}^\mathrm{ax} 
  = k_\mathrm{ax}\sqrt{\frac{\hbar}{2M\omega^\mathrm{ax}_m}}
  b_{km}^\mathrm{ax},
  \label{}
\end{equation}
\begin{equation}
  \tilde\eta_{km}^\mathrm{rad} 
  = k_\mathrm{rad}\sqrt{\frac{\hbar}{2M\omega^\mathrm{rad}_m}}
  b_{km}^\mathrm{rad},
  \label{}
\end{equation}
where $k_\mathrm{ax}$ ($k_\mathrm{rad}$) is the laser wavevector component
in axial (radial) direction. 
Only a part of the full Lamb-Dicke parameter matrix which corresponds 
to the illuminated ions enters the laser-ion Hamiltonian \eqref{eq:full_ham}. 
Let $(k_1, k_2)$ be two ions for which the MS gate is implemented. Then, 
we define the matrix $\eta_{im}$ used in the main text as
\begin{equation}
  \eta_{im} = \eta^\mathrm{rad}_{k_im}, \quad i = 1,2.
\end{equation}

\section{Pulse shaping with piecewise-polynomial pulses}
\label{appendix:spline_pulse_shaping}
As a first step of the pulse shaping procedure of the 
Section~\ref{subsec:pulse_shaping}, one needs to find 
$\Omega(t)$ which satisfies the Eqs.~\eqref{eq:gate_condition_alpha_nocar} and 
\eqref{eq:gate_condition_chi_nocar}.
These equations comprise a system of $2n_\mathrm{ions}$ linear equations 
and a single quadratic equation on $\Omega(t)$. The finite set of equations 
on a continuous function $\Omega(t)$ cannot define it uniquely, so additional 
constraints should be imposed on $\Omega(t)$. 

Assume that $\Omega(t)$ is decomposed into a basis set 
$\{b_s(t)\}$, $\Omega(t) = \sum_s \Omega_s b_s(t)$, where the functions $b_s(t)$ are 
defined on the interval $(t_0, t_f)$ of the gate duration. Then, the equations 
$\alpha_{im} = 0$ reduce to a linear system \cite{Wu2018}
\begin{equation}
  \sum_s  A_{ms} \Omega_s = 0,
  \label{eq:gate_condition_A_matrix}
\end{equation}
where 
\begin{equation}
  A_{ms} = \int_{t_i}^{t_f} dt\,b_s(t) \cos{(\mu t + \psi)} e^{i\omega_m t},
  \label{eq:A_matrix}
\end{equation}
and the equation $\chi_{12}(t_f) = \phi$ reduces to a quadratic equation
\begin{equation}
  \sum_{ss'} B_{ss'} \Omega_s \Omega_{s'} = \frac{\pi}{4},
\end{equation}
where
\begin{multline}
  B_{ss'} = -\sum_m2\eta_{1m} \eta_{2m}  \int_{t_i}^{t_f}dt\int_{t_i}^{t}dt'
  b_s(t) b_s(t')\times\\ 
    \times \cos{(\mu t + \psi)}\cos{(\mu t' + \psi)} \sin{\left[\omega_m(t-t')\right]}
\end{multline}

In order to ensure the smoothness conditions 
necessary for the consideration of Section~\ref{sec:pulse_shaping}, we 
search for $\Omega(t)$ in form of a cubic spline defined by its values in the points $t_s$ 
evenly spaced in the interval $(t_0, t_f)$, $t_s = t_0 + \frac{(t_f - t_0)s}{n_\mathrm{seg} + 1}$, 
where $s = 0,\dots n_\mathrm{seg} + 1$. The basis functions $b_s(t)$ are 
defined as cubic splines on the interval $(t_0, t_f)$ satisfying the  boundary conditions 
$b_s(t_0) = b_s(t_f) = 0$,  $b_s'(t_0) = b_s'(t_f) = 0$ and the conditions
$b_s(t_{s'}) = \delta_{s,s'}$, $s = 1\dots n_\mathrm{seg}$.


%

\end{document}